# The Aryabhata Algorithm Using Least Absolute Remainders

Sreeram Vuppala

## 1 Introduction

The year 2006 has seen renewed interest in the mathematics of Aryabhata (473- c.550), the great mathematician-astronomer of Classical India, for potential applications to cryptography. Rao and Yang [1] recently published an analysis of the Aryabhata algorithms for finding of multiplicative inverse of a group modulo a prime as well as the solution to multiple congruences. The RSA 2006 Conference held in San Jose in February also honored Aryabhata.

It is believed that Aryabhata was the *kulapati* (head) of the University at Nalanda in Magadha. He wrote at least two books: the *Āryabhatiya* and the *Āryabhata-siddhānta*, of which the latter is known only through references in other works. The *Āryabhatīya* presented Āryabhata's astronomical and mathematical theories, in which the earth was taken to be spinning on its axis and the periods of the planets were given with respect to the sun. The *Āryabhatīya* was translated into Arabic as *Arajbahara,* and in turn it influenced Western astronomers.

For general background to Indian mathematical tradition, see the books by Datta and Singh [2], Srinivasiengar [3], and Joseph [4]; and for its combinatoric and astronomical motivations see the recent papers by Kak [5-10]. Almeida *et al* [11] argue that Indian mathematics may have provided the spark that led to the scientific revolution in Europe, and this issue, together with a fresh appraisal of the larger question, is discussed by Pearce [12].

The *Āryabhatīya* presents a method to solve linear indeterminate equations, which may be adapted to find the multiplicative inverse in a group that is of interest in cryptography, signal processing, coding and computer design. Although Aryabhata calls this method the *Kuttaka* (the pulverizer), it is now better known as the Aryabhata Algorithm [13].

This paper presents an introduction to the Aryabhata algorithm. We do so by the use of the least absolute remainders, which can also be used to speed up the Extended Euclid's algorithm to obtain multiplicative inverses in a group. It may be noted that the Extended Euclid's algorithm is very similar to the Aryabhata algorithm, and historians believe that Aryabhata was the first to solve linear indeterminate equations.

The exposition of the Aryabhata algorithm provided here can have performance that could exceed what was described by Rao and Yang [1]. Section 2 presents the basic

Euclidean algorithm and the use of least absolute remainders as background. Section 4 and 5 present the Aryabhata algorithm with the least absolute remainders.

**2 GCD of two numbers using least absolute remainders**

Euclid's algorithm gives a very simple and efficient method for the determination of the greatest common divisor (g.c.d) of two numbers.

Let *a*, *b* be the two numbers whose g.c.d is to be computed. Since there is only the question of divisibility, there is no limitation in assuming that *a*, *b* are positive and $a \geq b$. We divide *a* by *b* with respect to the least positive remainder.

$$a = q_1 b + r_1, 0 \leq r_1 < b$$

Next we divide *b* by $r_1$,

$$b = q_2 r_1 + r_2, 0 \leq r_2 < r_1$$

Then, continue the process on $r_1$ and $r_2$.

$$a = q_1 b + r_1$$
$$b = q_2 r_1 + r_2$$
$$r_1 = q_3 r_2 + r_3$$
$$.$$
$$.$$
$$.$$
$$r_{n-1} = q_{n+1} r_n$$

Stop when $r_{n+1} = 0$. The g.c.d is the last non-vanishing remainder $r_n$.

Now remove the condition of the remainder $r_n$ being positive. The conclusions are the same except that $r_n$ may be a negative value of the g.c.d. It was shown by a German mathematician Kronecker (1823-1891) that no Euclid algorithm can be shorter than the one obtained by using the least absolute remainders [14].

We demonstrate these two methods using an example. Let $a = 76,084$ and $b = 63,020$.

Method 1:

$$76,084 = 63,020 \cdot 1 + 13,064$$
$$63,020 = 13,064 \cdot 4 + 10,764$$
$$13,064 = 10,764 \cdot 1 + 2,300$$
$$10,764 = 2,300 \cdot 4 + 1,564$$
$$2,300 = 1,564 \cdot 1 + 736$$
$$1,564 = 736 \cdot 2 + 92$$
$$736 = 92 \cdot 8$$

**Figure 1**

This method uses the Euclid's algorithm using least positive remainder.

Method 2:

| | |
|---|---|
| $76,084 = 63,020 \cdot 1 + 13,064$ | $76,084 = 63,020 \cdot 1 + 13,064$ |
| $63,020 = 13,064 \cdot 5 - 2,300$ | $63,020 = 13,064 \cdot 5 + (-2,300)$ |
| $13,064 = 2,300 \cdot 6 - 736$ | $13,064 = (-2,300) \cdot (-6) + (-736)$ |
| $2,300 = 736 \cdot 3 + 92$ | $2,300 = (-736) \cdot (-3) + 92$ |
| $736 = 92 \cdot 8$ | $736 = 92 \cdot 8$ |
| (a) | (b) |

**Figure 2**

The second method shows the Euclid's algorithm using least absolute remainders. In Figure 2, you see two representations of the equations. It can be easily seen [13] that the negative signs for the quotients can be dropped in the calculation of the g.c.d. In Figure 2.b, the representation keeps in mind of the negative signs of the quotients during the computations.

**3. Linear Indeterminate Equations and Multiplicative Inverses**

**Theorem 1:** When *a* and *b* are relatively prime, it is possible to find such other integers *x* and *y* that
$$a \cdot x + b \cdot y = 1 \tag{1}$$

In other words, unity is a linear combination of '*a*' and '*b*'.

Given positive pair wise prime integers '$a$' and '$b$', it is very often necessary to find the multiplicative inverses '$a^{-1} \mod b$' and '$b^{-1} \mod a$'. The inverses can be solved by solving equation (1).

3.1. Solving Linear Indeterminate Equations using Euclid's algorithm:

We suppose that $a > b$. To make the notation more systematic, we write $a = r_1$ and $b = r_2$ in stating the algorithm.

$$r_1 = q_1 r_2 + r_3$$
$$r_2 = q_2 r_3 + r_4$$
$$\ldots\ldots$$
$$r_{n-3} = q_{n-3} \cdot r_{n-2} + r_{n-1}$$
$$r_{n-2} = q_{n-2} \cdot r_{n-1} + 1$$

$$1 = -r_{n-3} \cdot q_{n-2} + (1 + q_{n-2} \cdot q_{n-3}) \cdot r_{n-2}$$
$$1 = r_{n-2} - q_{n-2} \cdot r_{n-1}$$

(a)          (b)

**Figure 3**

We shall obtain a representation in Equation (1) by a stepwise process derived from Figure 3(a). This is shown in Figure 3(b).

3.2. Extended Euclid Algorithm:

A straight-forward modification of the above method is the "The Extended Euclid Algorithm" (which is really a version of Aryabhata's Kuttaka algorithm that will be described formally later) that is used for the computation of multiplication inverses commonly used in cryptographic algorithms like the RSA encryption system. The multiplicative inverse is denoted as $a^{-1} \mod b$, where $a > b$.

The simpler version of Euclidean algorithm from [1] will illustrate the principle.

**Example 1:**
Let $a = 137$ and $b = 60$.

| $i$ | $r_i$ | $q_i$ | $x_i$ | $y_i$ |
|---|---|---|---|---|
| -1 | 137 | - | 1 | 0 |
| 0 | 60 | - | 0 | 1 |
| 1 | 17 | 2 | 1 | -2 |
| 2 | 9 | 3 | -3 | 7 |
| 3 | 8 | 1 | 4 | -9 |
| 4 | 1 | 1 | -7 | 16 |
| 5 | 0 | | | |

**Table 1**

$$q_i \leftarrow quotient(r_{i-2}/r_{i-1}), r_i \leftarrow r_{i-2} \bmod r_{i-1}$$
$$x_i \leftarrow x_{i-2} - q_i x_{i-1}, y_i \leftarrow y_{i-2} - q_i y_{i-1}$$

From the table 1, we have $x = -7 \bmod 60 = 53$ and $y = 16$. Therefore, $137^{-1} \bmod 60 = 53$ and $60^{-1} \bmod 137 = 16$.

3.3. The Improved Extended Euclid Algorithm:

This method uses the extended Euclid algorithm but uses least absolute remainders to solve for inverses. The list of quotients will use list obtained from Method 2(b) instead and method 2(a) and then use an iterative method to solve $x_i$ and $y_i$.

Using Example 1 in the previous section:

Let $a = 137$ and $b = 60$.

| $i$ | $r_i$ | $q_i$ | $x_i$ | $y_i$ |
|---|---|---|---|---|
| -1 | 137 | - | 1 | 0 |
| 0 | 60 | - | 0 | 1 |
| 1 | 17 | 2 | 1 | -2 |
| 2 | -8 | 4 | -4 | 9 |
| 3 | 1 | -2 | -7 | 16 |
| 4 | 0 | | | |

**Table 2**

$$q_i \leftarrow quotient(r_{i-2}/r_{i-1}), r_i \leftarrow r_{i-2} \bmod r_{i-1}$$
$$x_i \leftarrow x_{i-2} - q_i x_{i-1}, y_i \leftarrow y_{i-2} - q_i y_{i-1}$$

In some cases, the last remainder could be -1. If this is the case, then the initial conditions for $x_i = -1, 0$ and $y_i = 0, -1$.

Note: In Method 2, we can find the g.c.d of two numbers by neglecting the negative signs. In the case of multiplicative inverses, the negatives signs of the quotients should not be neglected.

**Example 2:**

We show another example to prove that the improved Extended Euclid's algorithm performs much when the least absolute remainders are used.

Let $a = 249$ and $b = 181$.

| $i$ | $r_i$ | $q_i$ | $x_i$ | $y_i$ |
|---|---|---|---|---|
| -1 | 249 | - | 1 | 0 |
| 0 | 181 | - | 0 | 1 |
| 1 | 68 | 1 | 1 | -1 |
| 2 | 45 | 2 | -2 | 3 |
| 3 | 23 | 1 | 3 | -4 |
| 4 | 22 | 1 | -5 | 7 |
| 5 | 1 | 1 | 8 | -11 |
| 5 | 0 | | | |

**Table 3**

This EEA took 5 steps to compute the inverses.

Using the improved Extended Euclid's algorithm:

| $i$ | $r_i$ | $q_i$ | $x_i$ | $y_i$ |
|---|---|---|---|---|
| -1 | 249 | - | -1 | 0 |
| 0 | 181 | - | 0 | -1 |
| 1 | 68 | 1 | -1 | 1 |
| 2 | -23 | 3 | 3 | -4 |
| 3 | -1 | -3 | 8 | -11 |
| 3 | 0 | | | |

**Table 4**

The improved algorithm solved the problem in 3 steps.

**4. Aryabhata Algorithm "Kuttaka"**

The algorithm written by Aryabhata as found in his book, the Aryabhatiya, is called "The Kuttaka" (the pulverizer). In [13], this method was introduced as the Aryabhata algorithm, to conform to the convention of associating the person with his result. This algorithm for the solution of a linear indeterminate equation appears to be the earliest recorded anywhere.

In the Aryabhata Algorithm, the list of quotients is called the "*Valli*". The second column in Table (5.a) is the "*Valli*".

The improved Aryabhata algorithm (IAA) will be discussed here for solving the multiplicative inverse $a^{-1} \mod b$ by solving the equation $a \cdot x + b \cdot y = 1$.

**Example 3:** Let $a = 137$ and $b = 60$

Using the recursive formula: $S_i = q_i \cdot S_{i+1} + S_{i+2}$ We start with $S_5 = 1, S_4 = q_4$

| $i$ | $q_i$ | $S_i$ |
|---|---|---|
| 1 | 2 | 16 |
| 2 | 3 | 7 |
| 3 | 1 | 2 |
| 4 | 1 | 1 |
| 5 |   | 1 |

**Table (5a)**

We can conclude that with $n = i - 1$, $y = (-1)^n S_1$ and $x = (-1)^n \cdot S_2$. Therefore, $x = -7$ and $y = 16$.

## 5. A Faster Improved Aryabhata Algorithm

The IAA can be further improved by reducing the "*Valli*" by using the least absolute remainders in Method 2. When the least absolute remainders method is used, the last remainder can end with 1 or -1.

Case '-1': $S_1 = S_1 \cdot (-1)$
Case '1': $S_2 = S_2 \cdot (-1)$
$$a^{-1} \mod b = S_2, \ b^{-1} \mod a = S_1$$

Using the same example:

| $i$ | $r_i$ | $q_i$ | $S_i$ |
|---|---|---|---|
| 1 | 17 | 2 | -16 |
| 2 | -8 | 4 | -7 |
| 3 | 1 | -2 | -2 |
| 4 |   |   | 1 |

**Table (5b)**

In this case, the last remainder is 1, so $S_2 = S_2 \cdot (-1) \Rightarrow (-7) \cdot (-1) = 7$
Therefore, $y = (-1)^n S_1$ and $x = (-1)^n \cdot S_2 \Rightarrow y = (-1)^3 \cdot S_1 = 16, x = (-1)^3 \cdot S_2 = -7$

**Example 4:** Let $a = 249$ and $b = 181$

| $i$ | $q_i$ | $S_i$ |
|---|---|---|
| 1 | 1 | 11 |
| 2 | 2 | 8 |
| 3 | 1 | 3 |
| 4 | 1 | 2 |
| 5 | 1 | 1 |
| 6 |   | 1 |

**Table (6a)**

Therefore, $x = -8$ and $y = 11$.

The faster method:

| $i$ | $r_i$ | $q_i$ | $S_i$ |
|---|---|---|---|
| 1 | 68 | 1 | -11 |
| 2 | -23 | 3 | -8 |
| 3 | -1 | -3 | -3 |
| 4 |    |    | 1 |

**Table (6b)**

Since the last remainder is -1, $S_1 = S_1 \cdot (-1) \Rightarrow (-11) \cdot (-1) = 11$

Therefore, $y = (-1)^n S_1$ and $x = (-1)^n \cdot S_2 \Rightarrow y = (-1)^3 \cdot S_1 = -11, x = (-1)^3 \cdot S_2 = 8$.

**Conclusions**

The use of least absolute remainders can improve both the standard Aryabhata algorithm and the extended Euclid algorithm in performance.

Rao and Yang stated [1] that the number of computations required to solve for multiplicative inverses is reduced using the Aryabhata algorithm, though the backdrop is that it needs extra memory to store all the quotients. With the use of the least absolute remainders, we have a best case scenario where the extra memory and computations is reduced by at least half. The worst case scenario using least absolute remainders will be the same as using the least positive remainders. Since, we know that multiplicative inverses are necessary in both the Chinese Remainder Theorem and the Aryabhata Remainder Theorem; the execution times can be greatly reduced when using the least absolute remainders.

Sreeram Vuppala
E-mail: svuppa1@lsu.edu